# ASTRA: Space-Based Mission for Ultrahigh Energy Particles

**Authors:** Mauricio Bustamante, Johannes Eser, Ke Fang, Claire Guépin, John Krizmanic[1], Eric Mayotte, Keith McBride, Kohta Murase, Mary Hall Reno, Frank Schroeder, Tonia M. Venters[2], Stephanie Wissel for PhysPAG

## 1. Science Investigation

Ultra-high–energy cosmic rays (ECR $\gtrsim$ 1 EeV) are the highest-energy particles known, signaling extreme particle processes at work in the universe. However, many aspects of their nature remain largely unknown, even after more than a century of study. Very-high-energy (Eν $\gtrsim$ 1 PeV) neutrinos associated with cosmic-ray interactions, both during the acceleration process and propagation, would provide new insight into these extreme particles, as we have seen at lower energies with the dawn of TeV neutrino astronomy. Nevertheless, only a handful of such neutrinos have been observed thus far. A space-based observatory dedicated to studying cosmic rays, neutrinos, and photons would provide an unprecedented platform for observations of these extreme-energy messengers.

Astro2020 [1] identified multimessenger astrophysics as a key priority area for the coming decades. A key discovery area for the *New Messengers and New Physics* theme is to "[Transform] our view of the universe by combining information from light, particles, and gravitational waves." Cosmic-ray and neutrino science were identified as required capabilities supporting this key discovery area. Cosmic rays and neutrinos also uniquely address Astro2020's key science questions: "What are the properties of dark matter and the dark sector?"; "Why do some compact objects eject material in nearly light-speed jets, and what is that material made of?"; and "Are TeV-PeV Neutrinos and ultra-high–energy cosmic rays produced in relativistic jets?".

### Key science gaps

<u>Where do the highest energy cosmic rays come from?</u> – Cosmic rays are subject to deflection by magnetic fields interleaving the cosmos between us and their sources. While our knowledge of these fields continues to improve through electromagnetic observations [2], the energies at which cosmic rays point back to their sources remains unknown. Indications of the extragalactic origins of cosmic rays have been established with the detection of a dipole above 8 EeV [3]. At higher energies, hints of hotspots [4] and correlations with nearby sources [5] have begun to emerge from the data. These observations provide encouraging signs that increasing full-sky exposure will bring the discovery of individual cosmic ray sources within reach.

1 Corresponding author

2 Corresponding author

What is the nature of extreme cosmic accelerators? – Evidence strongly suggests that cosmic rays gain their energy through cosmic particle acceleration. Viable source candidates must contain strongly magnetized regions that are large enough to confine cosmic rays until they reach ultra-high energies. Additionally, the source environment must provide the conditions to ultimately yield a favorable cosmic-ray composition mix [6] given likely energy losses within and outside of the source. Finally, the luminosities and source densities must reproduce cosmic-ray spectrum and composition measurements. In the face of these requirements, several candidate classes have been identified: newly born magnetars, gamma-ray bursts, tidal disruption events, jetted active galactic nuclei, and galaxy clusters [e.g., 2]. Charged particle astronomy, informed by composition measurements, will reveal whether these sources accelerate ultra-high–energy cosmic rays.

How are cosmic rays connected to gamma rays and neutrinos? – Astrophysical cosmic-ray accelerators tend to harbor intense matter and radiation fields with which the cosmic rays may interact, producing gamma rays and neutrinos. The strikingly similar intensities of the extragalactic gamma-ray background, the diffuse astrophysical neutrino flux, and the cosmic-ray spectrum suggests a common origin for these phenomena. Furthermore, IceCube has reported detections of individual sources of TeV-PeV neutrinos, providing crucial clues about PeV cosmic-ray acceleration [7, 8]. However, at higher energies, the diffuse astrophysical neutrino flux has yet to be revealed, and individual sources have yet to be identified. A substantial increase in exposure coupled with good angular resolution will broaden our search for the highest-energy neutrinos, which will in turn deepen our probe into the most extreme cosmic accelerators. Instantaneous coverage of large area of the sky will provide unprecedented opportunities for searching for very-high-energy neutrinos from transient phenomena [9,10].

What can the highest-energy particles teach us about particle physics at such extreme energies? – Searches for VHE and UHE ($E\nu \gtrsim 1$ EeV) neutrinos can probe new interactions, the presence of new particles and the nature of dark matter [11]. For example, indirect searches for super-heavy dark matter (SHDM) are sensitive to dark matter masses, annihilation cross sections and decays into neutrinos, with the potential for world-leading constraints for the highest masses and energies [12]. Hadronic interactions can be probed by UHE cosmic ray air showers where even above 40 EeV, a measurement of the proton-air cross section can be made [13].

**The Need for the ASTRA Incubator**

Community support is growing for a space-based cosmic ray and neutrino mission. The Astro2020 science questions have been further refined together with the national and international community through Snowmass whitepapers on VHE neutrinos [10] and ultra-high energy cosmic rays (UHECRs) [2], which detailed the unique benefits and opportunities offered by space-based observations.

A full-sky, high-aperture space-based experiment that measures UHECRs, UHE photons, and VHE neutrinos is essential to address the Astro2020 science questions, especially by having low systematic uncertainties in searching for common sources. Only a space-based mission can provide full-sky coverage with a single observatory, which is critical for two reasons. First, the astrophysical discovery potential of cosmic-ray and neutrino sources is driven by statistics at these energies. Second, this space-based mission reduces the systematic uncertainties compared to combining measurements of ground-based observations. Therefore, a space-based mission can best address the astrophysical goals of cosmic-ray physics by providing the highest exposure over the entire sky and with the required energy, angular, and mass composition resolutions. While the next-generation ground-based UHECR observatory is anticipated to more precisely probe UHECR composition evolution, the performance of fluorescence-only extensive air shower (EAS) measurements is expected to improve by incorporating Machine learning and using all data available in the profile [14], potentially exceeding the mass identification sensitivity needed to achieve moderate-resolution event-by-event rigidity reconstructions from a space-based observatory.

An earlier probe-class mission study for a space-based observatory that was based on the international JEM-EUSO Collaboration (POEMMA, [13]) studied the science case and a preliminary design. This has led to a series of pathfinder missions such as MiniEUSO [15] (in operation on board the ISS since 2019), EUSO-SPB2 [16] (flown in 2023), POEMMA Balloon with Radio [17] (PBR, planned flight 2028), and TERZINA aboard the NUSES satellite [18] (planned launch 2026). However, further incubation is needed to mature technologies needed for a space-based mission based on advances since the probe study:

(1) Enhanced launch vehicle capabilities would allow for different instrument architectures that could increase the cosmic-ray and neutrino exposures.
(2) PBR’s pioneering technique of combining optical Cherenkov and geomagnetic radio observations may improve the accuracy of air-shower measurements.
(3) More efficient propulsion could extend the mission lifetime.
(4) Improvements in slewing capabilities and refinements in orbital selection could enable faster follow up to transient alerts and optimize sky coverage.
(5) New approaches for rejecting noise events (e.g., bifocalizing optics) could enable lower-threshold observations, increasing event acceptance.

### 2. Science Table

| Science Objectives | Physical Parameters | Observables | Potential Challenges |
|---|---|---|---|
| *Determine the Astrophysical Source(s) of UHECRs and search for corresponding VHE neutrinos.* | *Flux, celestial sky distribution, and composition of UHECRs and neutrinos.* | *Fluorescence from EAS from UHECRs and neutrinos with good angular and $X_{max}$ resolution. Cherenkov EAS signal for VHE neutrinos.* | *Stereo observations needed. Requires good understanding of the inter- and extra-galactic magnetic fields.* |
| *Measure physics processes at the highest center-of-mass energies.* | *Nuclear cross sections at the highest COM energies.* | *Energy Spectrum, arrival directions, and composition measurements.* | *Resolution of nuclear composition measurements.* |
| *Search for Sources of Extreme Energy Neutrino Emission.* | *Measure the flux and celestial sky distribution of Ultra-high Energy Neutrinos.* | *Radiation from EAS induced by neutrinos, separated from those by UHECR EAS.* | *Understanding of the measurement requires good EAS profile discrimination between UHECR and neutrino induced events.* |
| | | *Energy Spectrum of UHE Neutrinos.* | |
| *Identify Energetic Transient Astrophysical Sources that emit VHE neutrinos using Target-of-Opportunity (ToO) observations.* | *Correlation in space and time of neutrinos with mutli-wavelength EM photons, gravitational waves.* | *Cherenkov signal measurements from EAS induced by neutrinos.* | *Requires slewable multi-$m^2$ area telescopes to image the source and co-aligned Cherekov signal at 10 ns sampling to us time scale.* |
| | *Energy Spectrum of VHE Neutrinos.* | | *Complete Sky Coverage for Transient Events.* |
| *Search for signatures of the decay or annihilation of Super-Heavy Dark Matter (SHDM).* | *Flux and celestial sky distribution of Neutrinos and Photons from SHDM Decay.* | *Fluorescence and Cherenkov measurements of UHE EASs induced by neutrinos and photons.* | *Observatory orbit needs to allow for large exposure for viewing the galactic center.* |

## 3. Instrument Description

The POEMMA probe mission employs two identical satellites flying in loose formation in 525 km altitude orbits that effectively uses the earth's atmosphere as a vast UHECR detector and the earth as a cosmic neutrino converter. Each POEMMA instrument incorporates a wide field-of-view (45°) Schmidt telescope with an optical collecting area of over 6 $m^2$ to measure the optical signals from the EAS from both UHECRs and VHE neutrinos. This is accomplished by using a hybrid focal surface of each telescope that includes two disparate sections. A larger area of the focal surface employs a fast (1 μs) near-ultraviolet camera using multi-anode PMTs for EAS fluorescence observations, from UHECRs and UHE neutrinos that interact in the earth's atmosphere. A smaller area of the focal surface incorporates an ultrafast (10 ns) optical camera using SiPMs for measurement of the beamed, Cherekov signal from upward-moving EAS induced by the earth-emergent leptons from VHE cosmic neutrino interactions in the earth. Each satellite has avionics with the ability to quickly slew to view a target-of-opportunity (ToO) when needed to optimize the detection of neutrino transients using the Cherenkov signal of the EAS viewed near the limb of the Earth.

## 4. Mission Implementation

- Dual manifest launch of the two identical spacecraft to LEO 525 km orbits at 28.5$^o$ inclination orbits with nominal 300 km spacecraft separation and pointed to view common atmospheric volume. Propulsion to maintain orbit and to reduce spacecraft separation for at least 5 per year long-duration neutrino transient observations and then reposition at 300 km separation for UHECR observations.
- 3 year mission requirement, 5 year goal.
- Spacecraft pointing, 0.1$^o$ control, 0.01$^o$ stability/knowledge, Slew rate: 90$^o$ in 8 min
- *Ways to leverage industry and commercial capabilities.*
    - The availability of SpaceX's Starship and the SLS have almost doubled the allowed diameter of a payload from 4.6 m to 8 m, which would allow a larger diameter telescopes that would lead to significant improvements in UHECR exposure and potentially VHE neutrino exposure.
    - Improvements in propulsion, such as green propulsion, provide longer orbit lifetimes leading to larger exposures and mission lifetime.
- *Ways to leverage the Artemis Program & infrastructure:* Rideshare for a mission like the Zettavolt Askaryan Polarimeter (ZAP) [19]: a swarm of lunar orbiting SmallSats that detect UHECR-induced Askaryan radiation from the lunar regolith.
- *Potential partnerships:* The JEM-EUSO international collaboration that has been co-developing missions including MiniEUSO [15], EUSO-SPB1 [20], EUSO-SPB2 [16] , PBR [17], the POEMMA probe-study [21], and the NUSES Collaboration that has been developing TERZINA [18].
- *Any cost saving initiatives.* None identified.

**6. Endorsements for this White paper : Name & Affiliation**

1. George Filippatos (University of Chicago)
2. Giuseppe Osteria (INFN Naples)
3. Brian F. Rauch (WUSTL)
4. Valentina Scotti (Unina & INFN, Italy)
5. Roberto Aloisio (Gran Sasso Science Institute)
6. Sonja Mayotte (Colorado School of Mines)
7. Rossella Caruso (University of Catania & INFN-CT, ITALY))
8. Soon-Wook Kim (KASI)
9. Haroon A. Qureshi (INFN Napoli)
10. Francis Halzen (UW-Madison)
11. Muhammad Mustapha Abdullahi (Gran Sasso Science Institute)
12. Matteo Battisti (INFN Roma Tor Vergata)
13. Lorenzo Perrone (UniSalento and INFN Lecce)
14. Etienne Parizot (APC, Univ. Paris Cité)
15. Athina Meli (NC A&T State University)
16. Wolfgang Zober (Washington University in St. Louis)
17. Caterina Trimarelli (Gran Sasso Science Institute)
18. Abhijit Roy (GSSI)
19. Kaliroe M W Pappas (Columbia U)
20. Alexander Novikov (University of Delaware)
21. Giulio Fontanella (Gran Sasso Science Institute)
22. Angela V. Olinto (Columbia U)
23. Najia Moureen Binte Amin (University of Delaware),

24. Michael Unger (KIT)
25. Toni Bertólez-Martínez (Wisconsin-Madison)
26. Paras Koundal (Bartol Research Institute, University of Delaware)
27. Luis Anchordoqui (Lehman College, City University of New York)
28. Foteini Oikonomou (Norwegian University of Science and Technology)
29. Sergio J. Sciutto (La Plata University)
30. Nathan M. Mellish (University of Utah)
31. Aswathi Balagopal V. (University of Delaware - Bartol)
32. Paula Gina Isar, (Institute of Space Science - INFLPR Subsidiary, Bucharest-Magurele, Romania)
33. Markus Cristinziani (Uni Siegen)
34. Ioana C. Maris (ULB)
35. Ivan De Mitri (GSSI)
36. Carlos J. Todero Peixoto (University of São Paulo)
37. Zbigniew Plebaniak (INFN Tor Vergata)
38. João R. T. de Mello Neto (UFRJ)
39. Serap Tilav (University of Delaware)
40. Yuca C. Chen (Bartol Research Institute)
41. Mario E. Bertaina (Univ. Torino, Italy)
42. Emmett Krupczak (MSU)
43. Arifa Khatee Zathul (UW Madison)
44. Paula N Gálvez Molina (University of Delaware)
45. Antonio Insolia - (Catania University and INFN - Italy)
46. Miles Garcia (University of Delaware)
47. Eva Maria Martins dos Santos (FZU - Institute of Physics of the Czech Academy of Sciences)
48. Jaime Alvarez-Muñiz (IGFAE, Univ. Santiago de Compostela)
49. Kathryn Plant (NRAO)
50. Marco Ricci (INFN Roma Tor Vergata and Frascati National Labs)
51. Kathryn Plant (NRAO)
52. Tsuguo Aramaki (Northeastern University)
53. Michael DuVernois (University of Wisconsin-Madison)